\documentclass{ws-ijqi}

\begin{document}

\markboth{Garry Bowen}
{Feedback in Quantum Communication}

%
\catchline{}{}{}{}{}
%

\title{FEEDBACK IN QUANTUM COMMUNICATION}

\author{GARRY BOWEN}

\address{Centre for Quantum Computation, DAMTP, University of Cambridge,
Wilberforce Road\\
Cambridge CB3 0WA,
United Kingdom\\
gab30@damtp.cam.ac.uk}

\maketitle

\begin{history}
\received{(Day Month Year)}
\revised{(Day Month Year)}
\end{history}

\begin{abstract}
In quantum communication feedback may be defined in a number of distinct ways.
An analysis of the effect feedback has on the rate information may be
communicated is given, and a number of results and conjectures are stated.
\end{abstract}


\section{Introduction}   

Communication is primarily the generation of correlations between two parties.
The transmission of information is undertaken to correlate the received message
with an possible set of transmitted messages.  In the classical theory of
information the maximum rate of information transmission through a memoryless
channel is given by a measure of the maximum correlation that may be generated
through the channel.  The correlation measure is the \textit{mutual
information} between the sender's alphabet and the receiver's output alphabet.
Utilizing feedback, where the receiver's output state is communicated
noiselessly to the sender cannot increase the maximum asymptotic rate of
transmission.  The capacity, defined as the maximum rate of information
transfer per channel use, cannot then be increased by the use of
feedback.\cite{cover}

Feedback in quantum information theory becomes a more slippery concept, mainly
due to the quantum no-cloning theorem.\cite{wootters82}  The output state of
the receiver cannot be copied with perfect fidelity, and so the natural
generalisation of the classical feedback scenario cannot take place.  A more
subtle approach is then needed in both the definition and implications of
feedback in quantum communication.

\section{Defining Feedback}

The implementation of classical feedback may be done independently of the
receiver.  The receiver may thus play either an active or passive role in the
feedback protocol.  In quantum communication the actual role of the receiver
can determine the type of feedback utilized.

\subsection{Active quantum feedback}

The natural extension to the classical feedback protocol is to assume that the
receiver may transmit an \textit{arbitrary} amount of information noiselessly
to the sender.  The feedback scenario then may be represented by a noiseless
channel from receiver to sender.  The receiver may process the output of the
channel in any way, and include arbitrary addition information to transmit back
to the sender.  This extension of the concept of feedback does not change the
fact that the capacity of the channel cannot be increased by feedback.  The
idea of a noiseless feedback channel is useful, however, because it may easily
be extended to the quantum case.  Active quantum feedback can therefore be
defined as \textit{the use of a noiseless quantum channel from the receiver to
sender}.  Again, the receiver is unrestricted as to the operations they may
perform before transmission through the feedback channel.

\subsection{Passive quantum feedback}

Passive quantum feedback is when the sender receivers information about the
output state without that information being communicated to the receiver.  Any
quantum channel may be modelled by a unitary transformation with an environment
in a known initial state.  From the Kraus representation\cite{kraus} of the channel, a
measurement of the environment will give a particular output state determined
by the outcome of the measurement on the environment.  Alternatively, the
entire environment state may be accessible by the sender\cite{harrow04} (labelled
\textit{coherent feedback} by Winter\cite{winter04}).

As shall be demonstrated later, the known upper bounds for active feedback
protocols also apply to any passive feedback protocol.  Any gains from passive
feedback protocols are thus necessarily less than those possible for known
optimal active feedback protocols.

\section{Capacities for Quantum Channels}

The Holevo--Schumacher--Westmoreland theorem\cite{holevo97,schumacher97} gives the achievable rates
for memoryless quantum channels.  This states that the maximum amount of mutual
information that may be generated through a quantum channel $\Lambda$ is
bounded by the maximum amount of ``classical'' correlation that may be shared
by states through the channel.  For $n$ uses of a channel, the upper bound
becomes the regularized term,
\begin{equation}
C = \lim_{n\rightarrow \infty} \max_{\rho_{RQ} \in \mathcal{D}} \frac{1}{n}
S(R:\Lambda^{\otimes n} Q)
\label{eqn:C_bound2}
\end{equation}
where the quantum mutual information is defined by $S(A:B) = S(\rho_A)+S(\rho_B)-S(\rho_{AB})$, for $S(\omega) = -\mathrm{Tr}\; \omega \log \omega$ the von Neumann entropy of the state.  The notation $\Lambda^{\otimes n} Q$ represents the effect of $n$ copies of the channel $\Lambda$ acting on the state $\rho_Q$.  The maximum is taken over all \textit{separable} quantum states $\mathcal{D}$.
Codes exist that can achieve any rate below $C$, with asymptotically
vanishing probability of error.

The entanglement--assisted classical capacity $C_E$ is the rate that classical
information may be transmitted through a memoryless channel when both the
sender and receiver share an unlimited amount of entanglement prior to
transmission.\cite{bennett99,bennett01a}  The entanglement--assisted capacity
is additive, and hence is given by the single shot expression,
\begin{equation}
C_E = \max_{\rho_{RQ}} S(R:\Lambda Q)
\label{eqn:C_E}
\end{equation}
where the maximization is over \textit{all} bipartite quantum states.  The
entanglement--assisted capacity may be strictly larger than the unassisted
capacity, an example being for the noiseless channel where $C_E = 2C$ from
dense coding.\cite{bennett92}

\section{Feedback and Entanglement}

The role entanglement plays in the utilization of a quantum feedback channel is
easily seen from the fact that an arbitrary amount of entanglement may be
shared between the two parties using the feedback channel.  The
capacity\footnote{The corresponding entanglement--assisted capacities for
quantum information are related to the classical capacities by a factor of one
half.  This is achieved by utilizing quantum teleportation and quantum dense
coding.} of the channel utilizing quantum feedback must therefore be at least
as large as the entanglement--assisted capacity $C^{\,\mathrm{QFB}} \geq C_E$.

When the two parties share prior shared entanglement, any classical feedback
may be used along with some shared entanglement to generate a noiseless quantum
channel via quantum teleportation.  Conversely, a noiseless quantum feedback
channel may be used to both share entanglement as well as feedback classical
information.  Therefore the two scenarios are equivalent, and the
entanglement--assisted capacity with classical feedback $C_E^{\,\mathrm{FB}}$
is equal to the capacity with quantum feedback $C^{\,\mathrm{QFB}}$.

It may be shown that any feedback protocol that utilizes a quantum feedback channel
has an additive upper bound given by the maximum conditional quantum mutual
information $S(R:\Lambda Q |R') = S(\rho_{RR'}) + S\big((\Lambda_Q \otimes \mathbb{I}_{R'}) \rho_{QR'}\big) - S(\rho_{R'}) - S\big( (\Lambda_Q \otimes \mathbb{I}_{RR'})\rho_{QRR'}\big)$, for any state $\rho_{QRR'}$ that is separable between $R$ and $QR'$.  By showing that this quantity is
necessarily less than the righthand side of (\ref{eqn:C_E}), the equality
$C^{\,\mathrm{QFB}} = C_E^{\,\mathrm{FB}} = C_E$ is obtained.\cite{bowen02a}
For a quantum feedback channel the analogy with the classical theory thus only
holds in the case of the entanglement--assisted capacity.

Examining the case of coherent passive feedback shows that the maximum
conditional quantum mutual information is also an upper bound.  This may be demonstrated in a similar way to the proof of the case of active feedback, with the feedback operations changed to reflect the use of passive feedback.  It is not known
whether the upper bound is tight in this case, although it appears unlikely to
be so.

\section{Classical Feedback and Quantum Channels}

When only classical feedback is allowed between the receiver and sender, no
entanglement may be shared through the feedback channel.  Indeed, it is known
that if the two parties do not generate entanglement through the channel then
the use of classical feedback cannot increase the capacity, and $C^{\,
\mathrm{FB}} = C$.\cite{bowen03a}  This criterion obviously includes all
product (or non-entangled) coding schemes as well as all entanglement--breaking
channels.

Recent work, however, has indicated that the ability to generate a higher rate of entanglement utilizing classical feedback may increase the classical capacity for some classes of quantum channels.\cite{bennett04}  The particular examples for which this has been shown are known as \textit{echo--correctable} channels.  At present, the results for echo--correctable channels rely on the assumption of additivity for the separation between the unassisted and classical feedback assisted capacities.

The relationship between classical feedback and \textit{quantum information} is
somewhat different.  The rate that quantum information, in the form of intact
quantum states or entanglement, may be transmitted is increased by the use of
classical feedback.  There may even be a separation between rates that unknown states may be transmitted and the rate that entanglement may be shared through some channels.

\section{Discussion}

As initially stated, communication is primarily the generation of correlations
between parties.  The use of feedback can increase correlations between the
communicating parties.  In proving the known cases it is shown that, with sufficient prior shared correlations,
\textit{feedback cannot increase the correlations between the senders message and the
receivers output state}.
Cases where feedback has a positive contribution to the capacity appear to rely on the increased ability to generate additional correlations, generally assumed to be in the form of shared entangled states.

The implications for classical feedback and classical communication are that
the necessary pre-existing correlations are inherent in the code that is
utilized in communication.  For a quantum feedback channel, entanglement may be
shared between the parties, however, if the necessary amount of entanglement is
pre-existing the rate cannot be increased further.  This relationship provides
evidence that the entanglement--assisted capacity is a natural generalization
for the quantum information domain.

\section*{Acknowledgements}

The author wishes to thank the organisers of FQI04.  This work was funded by
EPSRC grant numbers GR/S34090/01 and GR/S92816/01.

\end{document}